\documentclass[]{spie}  

\usepackage[]{graphicx}
\def\titlefont{\LARGE\bf} 
\title{Reply to comment on Y-00 quantum secure communication protocol} 

\author{
Osamu Hirota\supit{a,b,c}, 
\skiplinehalf
\supit{a}
Research Center for Quantum Information Science, Tamagawa University, 
Tamagawa-gakuen 6-1-1, Machida, Tokyo 194-8610, JAPAN\\
\supit{b}
21st century COE program, Chuo University, Tokyo, JAPAN\\
\supit{c}
CREST, Japan Science and Technology\\
}

\authorinfo{
Further author information: 
(Send correspondence to Osamu Hirota)\\
Osamu Hirota: address; 6-1-1, tamagawa-gakuen, machida-city, Tokyo, Japan.
E-mail, hirota@lab.tamagawa.ac.jp; 
Telephone, +81-42-739-8674\\  
}

\begin{document} 
  \maketitle 

\begin{abstract}
In 2000, an attractive new quantum cryptography was discovered by H.P.Yuen, which can realize secure communication with high speeds and at long distance by conventional optical devices.
Recently, a criticism of the Yuen protocol, so called Y-00, was made by Nishioka, and Imai group(Mitsubishi and University of Tokyo), 
and they claimed Y-00 is essentially a classical stream cipher. This paper shows that the claim is incorrect.
In particular, it is shown that the relation $l_i=r_i\oplus \tilde{k}_i$, which is their basis for attack, has no essential role for any security analysis. In addition, we give a brief introduction of the general logic for the security of Y-00 as direct encryption and also for key generation.

Several industries have started to make a test-bed of Y-00 for digital optical fiber highway, following Kumar's leading work. We hope that this discussion encourages experimental works which realize a secure communication against quantum computer and quantum attacks based on physical principle.
\end{abstract}



\section{INTRODUCTION}
A quantum key distribution or generation scheme  for  two legitimate users(Alice and Bob) is one of the most interesting subjects
in quantum information science, which was pioneered by C.Bennett and
G.Brassard in 1984[1]. In addition, several variations for the key generation protocol  have been proposed[2]. We emphasize that such results are great achievement and open a new science.  
Many researchers believe that the key distribution by single photon is on the verge of commercial application. However we should take into account the fact that the societies of electronics and communication, and of cryptography are basically not interested in the practical use of quantum cryptography based on single photon schemes. Although there is no means of solving such a serious argument,
we would like to  make the following comment.  The key distribution or generation is a very important, but it is very narrow sense that one defines quantum cryptography by  only  BB-84 and  similar principle.
Yuen, Kumar, and their group have pointed out that the quantum cryptography should involve other aspects, and called  quantum information scientist's attention to quantum cryptography based on another principle. 
It is clear that the single photon based BB-84  have serious performance limitations[3], such as possible communication distance and key rate, if the customer wants to use them at  modern digital communication network so called super digital highway.
The research like Gisin's work[4] to cope with such limitations and coherent state based BB-84 should be encouraged, and also  research like Northwestern University's group to investigate another scheme for achieving the same function should be welcome. So we would like to support both types of research.

The purpose of this paper is to make a comment on the recent paper of a criticism[5] to the Yuen protocol. In cryptography, trials of attack to proposed protocol are essential. So consideration of attack on Y-00 is welcome. But one should follow the basic principle of the proposed protocol, before to attack it.
In this sense, the Nishioka et al  attacks[5] basically is not grounded on an appropriate theory.  As a result, there is misunderstanding on the Yuen protocol. We will clarify their mistakes in this paper. To reader who ask for more fruitful results, we recommend  reference [6].

\section{Yuen protocol:Y-00}
\subsection{General representation of Y-00}
In 2000, the basis of the present Y-00 was proposed, which is a new kind of quantum cryptography based on a different principle[6, 7]. In 2002, several concrete implementations have been realized[8, 9]. In these papers it is called $\alpha\eta$ scheme. 
Let us  mention first what is  basic concept in their idea.
There are many  fundamental theorems in quantum information theory. The most important theorem for information processing of classical information by quantum states is the following: 
\\
\\
{\bf Theorem 1}: \\
{\it Signals with nonorthogonal states cannot be 
distinguished without error and  optimum lower bounds for error rate exist.}\\
\\
This means that if we assign nonorthogonal states for bit values 1 and 0,
then one cannot distinguish 1 and 0 without error.
When the error  probability is 1/2 based on quantum noise, there is no way to distinguish them, from quantum detection theory[10].
On the other hand, in the quantum case, one has to take a quantum attack into account. So one needs the well known quantum no-cloning theorem.
\\
\\
{\bf Theorem 2}: \\
{\it Nonorthogonal states cannot be cloned without error.}\\

A fundamental requirement of secure communication is, first,
to establish  that the channel between Alice and Eve is very noisy, but the channel of Alice and Bob is kept as a normal communication channel by physical structure. To realize it, Yuen employed a combination of  a  shared short  key  for the legitimate users and a kind of stream cipher with specific modulation scheme, following the above two theorems. We note that a main idea in his protocol is the explicit use of a shared short key and physical nature of scheme for cryptographic objective of secure communication and key generation. And also, one can say that the origin of security comes from receiver performance with versus without key. 

According to his paper, the application of Y-00 is, first, a direct encryption stream cipher in  conventional cryptography, and then to extend to key generation, but  not  one-time pad which is very inefficient. Here we emphasize that one should employ different security criteria for  direct encryption and  key generation.
For  direct encryption  with unavoidable channel noise, the criteria are given as follows.\\
\\
{\bf Criteria of security}: 
\begin{itemize}
\item[\rm(a)]  Ciphertext-only attack on data and on key: To get plaintext or key, Eve knows only the ciphertext from her measurement.
\item[\rm(b)] Known/chosen plaintext  attack: To get key, Eve  inserts  her known or chosen plaintext data into modulation system( for example, inserts all 0 sequence as text). Then Eve tries to determine key from input-output. Using the key, Eve can determine the data from the ciphertext.
\end{itemize}
For key generation, unconditional security is preferable.
 
Here let us introduce a structure of Y-00. The data bit is modulated by $M$-ary keying driven by pseudo random number generator. 
The $M$-ary keying has $M$ different basis based on 2$M$ coherent states. So the data bit is mapped into one of 2$M$ coherent states randomly. In general, the quantum information system is described by a density operator. The density operator of the output of the coding/modulation system of Y-00 as seen by the attack for ciphertext-only individual attack is 
\begin{equation}
\rho_{T}=p_0\rho_0 + p_1\rho_1
\end{equation}
where
\begin{eqnarray}
\rho_0 &=& \sum q_j|\alpha_j\rangle\langle \alpha_j| \\
\rho_1 &=& \sum q_k|\alpha_k\rangle\langle \alpha_k|
\end{eqnarray}
The probability $p_i$ depends on the statistics of the data, and $q_j$, $q_k$ depend on the pseudo random number with $j$, and  $k$ being even and odd number, for example. Eve has to extract the  data from the quantum system with such a density operator.  However, according to  one of the most fundamental theorem(theorem 1) in quantum information theory, the accuracy of Eve's measurement is limited.
For  ciphertext only attack, the best way of Eve is of course given by the quantum optimum detection for two mixed states :$\rho_0$ and $\rho_1$. That is, the accuracy of measurement of Eve is as follows:
\begin{equation}
P_e= \min_{\Pi}(p_1Tr\rho_1\Pi_0 + p_0Tr\rho_0\Pi_1)
\end{equation}
As shown in references [8,9], the error probability of Eve as determined by the quantum limit is $\sim 1/2$ from the appropriate choice of the number $M$ and signal energy. It means that Eve's data is completely inaccurate.
In addition, {\it a Overlap Selection Keying}:{\bf  OSK} was proposed in references [11,12], based on the discussion with Yuen. Each set of basis state is
used for $\{1, 0\}$, and $\{0, 1\}$, depending on sub running key.
\begin{eqnarray}
Set\quad A_1 : 0 &\rightarrow& |\alpha_{(1)}\rangle, 
\quad 1 \rightarrow |\alpha_{(M+1)}\rangle \nonumber \\
Set\quad A_2 : 0 &\rightarrow& |\alpha_{(M+1)}\rangle, 
\quad 1 \rightarrow|\alpha_{(1)}\rangle \nonumber
\end{eqnarray}
The density operators of 1 and 0 for Eve are $\rho_1=\rho_0$. So Eve cannot completely estimate information bits.
This is an advantage of the OSK.
For  known/chosen plaintext  attack, Eve knows the data. So the best way for Eve is to detect 2$M$ pure coherent states which convey directly running key sequence. In this case, the accuracy of the data is also given by the quantum detection for 2$M$ pure coherent states[10,11,12] and the data involve unavoidable error given by 
\begin{equation}
{P}_e = \min_{\Pi} (1 - \sum p_iTr \rho_i\Pi_i)
\end{equation}
For appropriate $M$ and signal energy, we have $P_e \sim 1$. 
As a result, Eve's data involves unavoidable error even in any attack schemes, and she cannot get any meaningful information.

Thus, Y-00 is a new type of quantum cryptography based on quantum detection theory. That is, the security is guaranteed by quantum noise, but the system can be implemented by  conventional devices. {\it As quantum advantage,  Y-00  provides secrecy against  quantum computer and quantum attacks}. We can summarize the property of Y-00 as follows:

{\bf Direct encryption}\\
In the cipher-text only attack, Y-00 may exceed the classical Shannon limit. That is,
\begin{equation}
H(X|Y_E) > H(K)
\end{equation}
where $X$ is information, $Y_E$ is ciphertext as "measured value" for Eve, and $K$ is initial seed key. Even with $H(K|Y_E, X)=0$,  intuitively, the search complexity is between $2^{|K|}$and $ 2^{2^{|K|}}$ in Y-00, $|K| \sim 100$. For known plaintext attack, it will be expected that 
\begin{equation}
H(K|Y_E, X) > 0
\end{equation}
which corresponds to {\bf information theoretic security}.

As a result, Y-00 may provides information-theoretically secure scheme with very efficient performance for direct encryption.

{\bf Key generation}\\
The condition for secure key generation is 
\begin{equation}
H(X|Y_E, K) > H(X|Y_B)
\end{equation}
where $Y_B$ is Bob's observation with knowledge of the seed key. Recently it has been claimed that the key generation scheme using Y-00 with several randomizations can be {\bf unconditionally secure}[6]. 

In order to realize the above performance, several new randomizations will be introduced in reference [6].

\subsection{Classical Y-00}
We would like to show additional property of Y-00 here. The structure of Yuen protocol is formed by physical processes with specific modulator performance and so on. This means that the protocol is different with conventional cryptography formulated by mathematical concept, even the scheme is constructed by devices based on classical physics. It  is called "physical cryptography". Since the protocol is constructed by combination of physical processes, one can devise a new randomization that is possible with physics to increase the security. As a result, Y-00 may have better performance than conventional cryptography even the system is classical one. That is, to our surprise, this is a new type of stream cipher in the classical limit even in conventional cryptography.
Then the security of Y-00 is enhanced by classical noise and also quantum nature of the system. These facts are essential in order to understand Y-00.

\section{The attack on Y-00}
Recently Nishioka et al[5]have given an attack scheme on Y-00 as key generation scheme based on a relation of $M$-ary coding. But the original paper[8,9] of Y-00 is of direct encryption. Thus, although it involves many misunderstanding, in this section we try to use their scheme by the correct way in order to check whether it has a sense as an attack on Y-00. Then we will point out that the proposed attack scheme is meaningless even if the system is a completely noiseless classical system, and also basis(Eq(9)) of their claim has no essential role for Y-00.

\subsection{Noiseless classical system}
Their logic is that one can neglect the noise effect by choosing appropriate indirect observable for the data bit, and it reduces to less noisy or classical noiseless model.Then by combination of one-time pad for classical communication process, the security of the total system is equivalent to conventional stream cipher.
It seems that their claim is that {\it Y-00 corresponds to a classical cryptosystem, because $H(X|Y_E)\le H(K)$ up to $H(K)$ which means that there is no possibility of key generation, i.e, obtaining a new key from $X$ statistically independent of $K$,  even when the data is information-theoretically secure}.

Although their logic is not grounded on an appropriate theory, here we can verify their misunderstand.
Let us show their scheme. In order to clarify the essential point, we employ  one mode classical $M$-ary phase shift keying signals. That is, we do not need quantum signals(coherent state and so on).
Here we use their notations. The attack scheme to classical Y-00 can be summarized as follows:
The phase sift keying scheme used in Y-00 of reference [8] is taken to be
\begin{equation}
l_i=r_i\oplus \tilde{k}_i
\end{equation}
on the phase space, where $l_i$ is one of two regions separated by appropriate basis on the phase space.

If the appropriate axis is horizontal axis, $l_0$ is upper plain, $l_1$ is down plain. $r_i$ is true random data bit. $\tilde{k}_i$ is 0 for even number and 1 for odd number in the running key of $M$-ary assignment.
Eq(9) indeed holds as the coding scheme in the original scheme. 
For example, 
\begin{eqnarray}
(l_i=up,\quad \tilde{k}_i=even) &\longrightarrow& r=1 \nonumber \\
(up,\quad odd) &\longrightarrow& r=0 \nonumber \\
(down, \quad even) &\longrightarrow& r=0 \nonumber \\
(down,\quad odd )&\longrightarrow& r=1 \nonumber 
\end{eqnarray}
Let us define the sequences of numbers $l$, $r$, $\tilde{k}$ as follows:
\begin{eqnarray}
L&=&(l_1, l_2, l_3, \dots) \\
R&=&(r_1, r_2, r_3, \dots)\\
\tilde{K}&=&(\tilde{k}_{1}, \tilde{k}_{2}, \tilde{k}_{3}, \dots)
\end{eqnarray}

First we analyze direct encryption based on their scheme.
Let $K$, and $N$  be an initial key with length $|K|$, and length $|N|$ of pseudo random number, respectively.
The essential point of their attack is to measure indirect observable $L$. However, since the observable does not contain the information of the data bit, they are asked to use the Brute force attack for key to find a correct sequence of the data.
Here we can define that  $\cal{R_A}$ is a set of data random number sequence with the length of $|N|$.
Alice sends a sequence  $R_T$ in $\cal{R_A}$, and it is coded based on Eq(9) with a $\tilde{K}_j$, $j \in 2^{|K|}$. 
The proposed attack requires the measurement of $L$. Here  $l_i$ of $L$ depends on the axis selected on phase space at the first step. Here, let the horizontal axis be the selected axis. So $l_{i}$ is up or down. $\{C_{+},C_{-}\}$ in the attack scheme in their paper correspond to  $\{even$ and  $odd\}$.
 $\tilde{K}_j$ corresponds to pseudo random number sequence which  has a number of possibilities of $2^{|K|}$ and the length $|N|$.
Let  $R_T$, $L_T$, $\tilde{K}_T$ be true sequences used and defined on the phase space for Alice and Bob.
Let $L_m$ be the measurement result. She tries to assign  all kind of $\tilde{K}_j$ to her measured sequence $L_m$. So she gets 
a set $\cal{R}_{\rm{E}}$ based on $l_i=r_i\oplus \tilde{k}_i$. If $L_m$ is error free, then it is guaranteed that one of $\cal{R}_{\rm{E}}$ is the true random bits sequence. At this stage, we can say $H(X|Y_E)=H(K)$. So it is information-theoretically secure.
Here, if there is one bit error in $L_m$ by some reasons, then Eve has $L_T\oplus \bf{e}$, where ${\bf{e}}=(0,0,1,0,0, \dots)$ is error sequence. The position of the error is unknown and uniformly distributed. When Eve applies $\tilde{K}_j$, $j \in 2^{|K|}$ to $L_T\oplus \bf{e}$, then  it is not guaranteed that the true $R_T$ exists in $\cal{R}_{\rm{E}}$.
Since Eve does not know the true random bits sequence, she has to try $2^{2|K|}$ greater than the initial one in the sense of exponential. If there are many error, then it becomes 
$\sim 2^{2^{|K|}}$. So one may obtain Eq(6).

Since Y-00 is a physical cryptography, we should clarify the physical nature.
Let us discuss a property of physical system of $M$-ary PSK coding scheme as a physical cryptography. In general, phase spaces of Alice-Bob and Alice-Eve are not  same. The phase space is formed by the relative phase based on local phases of Bob and Eve. For example, quadrature amplitudes are  $\{x_c=A\cos(\phi_S-\phi_{L(Bob)}), x_s=A\sin(\phi_S-\phi_{L(Bob)})\}$, 
$\{x_c=A\cos(\phi_S-\phi_{L(Eve)}), x_s=A\sin(\phi_S-\phi_{L(Eve)})\}$. The Eq(9) is defined for the phase space of Alice-Bob. They assumed that Eve has the same phase space with that of Alice-Bob. In general Eve does not know the correct phase space. Because, the channel for Alice and Bob has synchronization, but the channel between Alice and Eve does not have the same synchronization. This fact is one of characters of physical cryptography. So Eve never understands what is the  axis decided by herself, and she never knows what kinds of $\{C_-, C_+\}$ should be used.
As a result, $\cal{R}_{\rm{E}}$ involves many errors, even when the measurement itself of $l_i$ is noiseless. Since obtained data  are random number,
she has no way to know which bits are incorrect. If the period of the pseudo random number is $2^{|K|}$ as in the case of a maximum length linear feedback shift register, then the number of possibilities is $\sim 2^{2^{|K|}}$ which is greater than $2^{|K|}$. 

Also the above physical situation provides information-theoretic security for known plaintext attack based on their attack. Of course, if one uses DSR(deliberate signal randomization) discussed in the next subsection, then more effectively the security is proved.

For the key generation, the original paper did not claim. But Nishioka et al made a new model as follows:
Eve does not assign $\tilde{K}_j$, $j \in 2^{|K|}$ at the first stage, and they forge a new model so called one-time pad scheme used Y-00 which was not totally mentioned in the original paper.
In their model,  they use the relations as follows:
\begin{eqnarray}
c_i&=&x_i \oplus r_i \\
c_i \oplus l_i &=& x_i \oplus \tilde{k}_i
\end{eqnarray}
where  $c_i$ is ciphertext, $x_i$ is plaintext.
The random number $R$ as the key disappears.
Eve can get $c_i$ and $l_i$, but she does not know $\tilde{k}_i$.
At this second stage,  Eve  tries  to assign to the Eq(9) all the different running keys in which 
the number of possibilities is $2^{|K|}$ . As a result, they claim that the security of Y-00 is equivalent to that of classical stream cipher $C=P\oplus K_{run}$, where $K_{run}$ is the output of the pseudo random number generator in stream cipher. 
However, this scheme does not work by the same reason as that for the first stage.

For a general concept of quantum key generation, we will discuss in the section 4.

\subsection{Randomization}
In general, one uses several randomizaions in conventional cryptography. We have randomizaion for the protection of Y-00 for the above attack, even the system is noiseless. The DSR(deliberate signal randomization) is one of the most effective method.
Although it is easy to show that the initial key advantage hides the basic axis of phase space(refer communication theory and [13]),  here we dare to take a perfect synchronization of Alice, Bob, and Eve at the initial stage of the protocol. Then the system is completely noiseless classical one. Since they insist that the system is regarded as classical when they measure indirect observable, they can use Eq(9).
However, for such an attack scheme, we can use 
\begin{equation}
L \oplus F = R \oplus \tilde{K}
\end{equation}
where $F$ is forced randomization which destroy the phase synchronization. This is one of DSR explained in the reference [6]. Such a randomizations induces error for the measurement of $L_m$ and destroy the relation like Eq(9).
So Eve's access information from the measurement: $L_m$ is completely zero. That is, Eve's trial becomes from $\tilde{K}_j, 
j\in 2^{|K|}$ to $2^{2^{|K|}}$ which means search trial for each bit in the sequence of length $2^{|K|}$. 

If Eve has only attack such as Eq(9), then this classical Y-00  for ciphertext only attack is already secure in the sense of Eq(6)  and it makes secure scheme(Eq(8)) for known plaintext attack even in the noiseless classical framework. So no quantum computer can break this scheme. 
As a result one can see that the attack used Eq(9) has no gain even in the noiseless classical Y-00. Again, this means that if Eve has only such an attack, then we do not need a quantum scheme to achieve "our purpose". 

\subsection{Effect of quantum noise in up-down measurement}
In the original experiment by Kumar's group, they did not use the randomization for simplicity. Even so, it will be secure for the above mentioned attack.
In the framework of their claim, let us show that the error of the measurement for $l_i$ is unavoidable,
even if they have complete synchronization for phase space.
The density operators of signal sets for up and down measurement are 
\begin{equation}
\rho_{up}=\sum \frac{1}{M}|\alpha_{up}\rangle \langle \alpha_{up}|, 
\quad
\rho_{down}=\sum \frac{1}{M}|\alpha_{down}\rangle \langle \alpha_{down}|
\end{equation}
We calculated the quantum  limit, which is the most rigorous lower bound of error probability, for this signal when the coherent state is mesoscopic and several hundreds of $M$. As a result, the error is several percents: $P_e \sim 0.01$. This means that the number of error bits is $P_e \times 2^{|K|} \gg 1$ which is enough for protect against the proposed attack with Brute force attack.
In addition, in the system of  Northwestern University, $M$ is over 2000. So if one does not know the running key, the measurement of $L$ would involve errors for a bundle of states close to selected axis by Eve. Since the basis state of Alice is chosen randomly and uniformly on the circle of phase space, the error bits of the measurement sequence $L_m$ will be distributed uniformly. The security  based on search is strongly enhanced even in their model. Thus the attack based on indirect measurement cannot overcome the security for direct measurement based on quantum detection theory[10]. However, before we discuss quantum effect, rather their attack is broken by classical randomization et al..

\section{Quantum Y-00 with information-theoretic security}
Let us remind that if Eve wants to make a situation
\begin{equation}
H(X|Y_E) \le  H(K)
\end{equation}
Eve has to get complete  information for the sequence by measurements, not for $l_i$. Nishioka et al do not mention the concrete measurement process they use. This is very strange and their second mistake. If they want to use compound relative of the measurement for the data and Eq(9), then they should show how to measure it. For any cases, the accuracy of Eve's measurement is bounded by quantum detection theory.
In the previous section, we focused on only the attack scheme given by Nishioka et al. 
There is more general description of the security of Y-00 independent of any concrete attacks, based on quantum detection theory. In the following, we introduce the general one.

If Eve wants to know some information on the data bits, she has to measure the signals by any instrument. In  Y-00 of the original model, the problem of the measurement ability reduces to the comparison with optimum binary quantum measurement and optimum phase measurement. Since Eve does not know $K$, she needs to make the phase measurement in order to identify $X$ for all possible basis selection from the running key.
According to the quantum detection theory, when Eve and Bob have the ultimate ability(ultimate receiver devices, and so on), their error probabilities may be shown that[6]
\begin{equation}
{P_e}^B \sim \exp(-4S) \quad vs \quad {P_e}^E \sim \exp(-2S)
\end{equation}
where $S=<n>$ is signal energy. Thus the error probability of Bob is smaller than that of Eve. This fact gives an advantage distillation, so it leads to information-theoretic security and key generation even one does not need to use any randomizations. That is, Eq(8) holds.

On the other hand, the randomizations provide more efficient scheme, as an example,
\begin{equation}
{P_e}^B \sim \exp(-2S') \quad vs \quad {P_e}^E \sim \exp(-S')
\end{equation}
where $S'\sim 10$ or reasonable value with DSR for $S=<n> \gg 1$.
The above explanations, in principle, beat off a comment such that Y-00 corresponds to classical cryptography.

\section{Conclusion}
We clarified that the criticism of Yuen protocol by T.Nishioka, T.Hasegawa, H.Ishizuka, K.Imafuku, and H.Imai is wrong, and the attack does not say anything for Y-00. However, the security of Y-00 is a different problem. But one can see reference [6] and subsequent papers for general theory. The author's group is concerned with how to realize Y-00 with ultimate security by intensity modulation/direct detection optical communication system as discussed in [11,12].

\section*{Acknowledgment}
OH is grateful to H.P.Yuen, P.Kumar and many colleagues of Northwestern University and Tamagawa University for discussions.


\end{document}